# Towards an Accessible and Rapidly Trainable Rhythm Sequencer Using a Generative Stacked Autoencoder

**Alex Wastnidge**
Department of Musicology
University of Oslo
alexanjw@uio.no

## ABSTRACT

Neural networks and deep learning are often deployed for the sake of the most comprehensive music generation with as little involvement as possible from the human musician. Implementations in aid of, or being a tool for, music practitioners are sparse.

This paper proposes the integration of generative "stacked autoencoder" structures for rhythm generation, within a conventional melodic step-sequencer. It further aims to work towards its implementation being accessible to the average electronic music practitioner.

Several model architectures have been trained and tested for their creative potential. While the currently implementations do display limitations, they do represent viable creative solutions for music practitioners.

## 1. INTRODUCTION

As outlined by Briot [2] and Yin[19], there is a notable history of music generation using neural networks and deep learning techniques. A trend noted here though, is the focus on expanding the capabilities of the artificial intelligence systems in the name of being able to perform more and more complete automatic music generation (AMG). Systems such as DeepBach[10], DeepHear[14], MusicVAE[12], Google Magenta[6] and its associated DAW plugins, as well as more examples present in [3][4], represent either fully-automated music generation, or the generation of musical accompaniment, generally in the style of an established musician or composer. For computer scientists, this understandably represents notable forward progress in the field of artificial intelligence. In these scenarios, however little to no consideration is often given to the needs and wants of the human music practitioner.

Two further key barriers exist for musicians and artists to engage with artificial intelligence as part of their practice. As outlined by Fiebrink [8], firstly, is that large datasets are often required to effectively train these systems. Secondly is that the computational load required is either inaccessible to the lay-user or is such that real or near-real-time processing is impractical.

This paper proposes and works towards a collection of tools and processes aiming to bring greater access and usability of generative AI to the contemporary music practitioner. This aims to address the three concerns outlined above with the goals of:

1. Creating a generative AI system which specialises in the rhythmic aspects of a melodic musical part, leaving the remaining creative work to the human agent.

2. Developing a dataset pipeline whereby a model can be trained on a collection of audio loops.

3. Implementing the system in a manner whereby the training and inference can occur quickly and on a consumer machine.

The implementation present here represents an expansion on the idea of the conventional step sequencer, which has been present in electronic music since the early days of the modular synthesiser. Via user input, the generative model contributes the rhythmic information of the sequence with pitch and other sequencer controls retained by the human agent. Pure Data is used for the user's musical interface, Keras Tensorflow written in Python code is used for all machine learning processes with Open Sound Control being used to relay inputs and outputs between the two.

## 2. RELATED MATERIAL

**2.1 Algorithmic, Probabilistic and Stochastic Processes and "Generative Music"**

Generative music involves the design of custom music systems which can generate musical parts or elements, based on rules reflecting the musical sensibilities of the user. This can involve, for instance, algorithmic [17], probabilistic or stochastic processes. Importantly, this practice balances the involvement of machine processes, with human interaction and curation, sensibilities which have informed the work here.

Noted also by Briot [2, 3, 4] as techniques tangental to those in machine learning are stochastic probabilistic processes such as Markov chains. Interestingly, as noted by Yin[19] when compared to deep learning networks, these perform equally well at AMG in human listening tests. They do however, lack much of the user intractability necessary for this implementation.

## 2.2 Accompaniment Tools

While tools for musical accompaniment are not the focus of this project, there are techniques used from elements of their implementation. These systems, noted in [2], have mostly grown from the work of Todd[15], acting as a precursor to more established AGM systems using Recurrent Neural Networks (RNNs), Long-Short Term Memory (LSTMs)[11], as well as projects such as MusicVAE[12] and Magenta[6]. The success of these systems is in their ability to recognise and generate long term patterns[19], which is not needed for this implementation.

MiniBach[2, 3] and DeepHear[14] utilise autoencoder structures and are trained on symbolic representation of music. As seen in those examples and expanded upon by Briot [3], both use a version of "one/many-hot" encoding to represent time-slices of MIDI piano roll data. This process is similar to how data was fed to the model here. In the case of DeepHear, seed data is fed into the decoder stage to generate new data, which has also been used here.

## 2.3 Creative Machine Learning Uses and Retooling

Beyond the realms of AMG, are tools for creatives to make use of machine learning algorithms. Notably so is Fiebrink's[7] Wekinator, which offers an accessible user interface for the training and use of classification and regression. Examples of its use often revolve around interactive music systems design, allowing for easier mapping of gestures and continuous values to musical or sounding parameters.

Cesar et al.[5] detail examples of uses and "re-tooling" of both Machine Learning and Music Information Retrieval (MIR) techniques for live electronic music performance. As well as creative mapping, this cites using the techniques in data-driven compositional audio programming. In these cases, "successful' machine learning technique is often secondary to its creative usage.

Fibrink[8] further details the various applications and creative uses of machine learning, amongst them "collaboration", though by her own admission, this is an "under recognised strength." The sensibilities of Fiebrink's approach to empowering musicians' use of machine learning and Cesar et al. use for creative ends has been of notable influence here.

## 2.4 Towards Generative AI Music Tools

Following on from this, examples of generative AI tools and techniques for creativity[18] have emerged. Google Magenta[6] has released DAW plugins which allow for generation of rhythmic and melodic parts but offer limited user controls and only pre-trained models.

Privato et al. use a combination of LSTM and Markov chains in their application "Scramble" which also offers an interface for users to train the model on their own MIDI datasets. Scramble does however, skew towards the realms of AMG offering full polyphonic compositional output, albeit based on the training data provided by the user.

Tokui's work with M4L.RhythmVAE[16] is of huge relevance here. M4L.RhythmVAE also has an accessible user interface for training and, like the intention here, its goal is rhythm generation. The implementation is as a Max for Live device, making it possible to integrate into the workflow of many electronic musicians.

The implementation here differs from Tokui in some key ways, however. Firstly, M4L.RhythmVAE can only be trained on MIDI data, the challenges of which are explored in Section 4. Secondly, as the name suggests, M4L.RhythmVAE utilises a variational autoencoder, whereas a more conventional autoencoder structure is used here. Thirdly, M4L.RhythmVAE aims to fully generate drum parts based on the training data, which is not the focus here.

## 3. SYSTEM DESIGN

### 3.1. Stacked Autoencoders

The implementation here is based on the principals and techniques outlined by Briot in [3], with the use of a "stacked autoencoder" architecture for 'Ex Nihilo' generation. The term "stacked autoencoder" refers to the idea to "hierarchically nest successive autoencoders with decreasing numbers of hidden layer units". According to Briot[3], by sending the data through several layers of compression, more and more high-level features are extracted, resulting in a latent layer with a high-level encoded form. After training, by isolating only the decoder phase, new values can be fed by the user into the latent layer which are then fed-forward through the decoder to generate new data.

The appeal of this technique is that it offers a balance between exploiting the capabilities of generative neural networks while retaining avenues for human control in-line with the principals of generative music practice.

### 3.2. Input Encoding

Also used from Briot[3] as well as [14], are the techniques for feeding symbolic musical representation into the system using "many-hot" encoding. For reasons explained in Section 4, audio loops were used as the training data but were subsequently encoded to this "many-hot" format.

Using the Librosa library in Python, an audio loop is loaded in and has its onsets detected. These represent each rhythmic event present in the loop. With the onsets detected, using the pretty_midi library these are converted from a seconds time-base to a MIDI-ticks time-base and used to create a new MIDI file. MIDI-ticks allowed for suitable resolution for the imperfect, unquantizied audio loops. The created MIDI file's note information is then looped through, rounding each tick onset value to the nearest sixteenth-note tick. This is then used to create a "many-hot" encoded one-dimensional array with a length of thirty-two, representing the thirty-two sixteenth-note time steps in a two bar sequence.

Where this implementation differs from the aforementioned [2, 14], is that rather than feeding sixteenth-note time-slices from piano roll representation, the entire two bar sequence is fed to the model with each time step as a separate input.

### 3.3. Architecture

With the "many-hot" encoding established, the input and output size of the model is always thirty-two. Beyond this, several architectures for the autoencoder were trialled to explore balances between best machine learning practice and most engaging creative results.

As binary data was used as input, binary cross-entropy was used to measure the loss function[2, 9]. With the exception of the output layer, the activation function used throughout was the *rectified linear unit* (ReLU)[1, 2]. This combination allowed for the binary data to be encoded and decoded in a continuous manner[1], allowing for greater variance in the latent space when fed with binary data.

At the output, the sigmoid function was used in order to receive values only between zero and one[1, 2]. As in [14] the output values are thresholded in order to create the necessary binary values. Initially this value was set to 0.5, as in [14] but this value was later opened to user control.

In the first instance, an architecture using a single encoder/decoder layer of sixteen and a latent layer of two was used (Figure 1). This represented the smallest possible architecture while still qualifying as a "stacked" autoencoder.

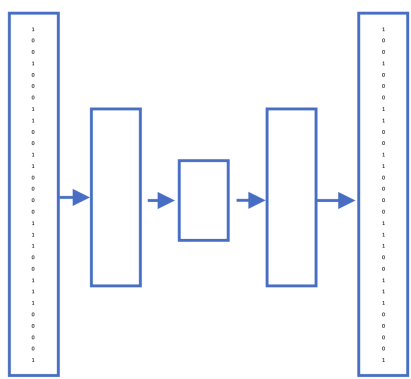

32 input -> 16 encoding -> 2 latent -> 16 decoding -> 32 output

**Figure 1: Initial autoencoder architecture with example binary data**

Once primary functionality had been achieved, three further models were trained and tested. The first was a simple expansion of the stacked autoencoder adding an additional hidden layer for encoding/decoding.

The second used the same expanded structure but used weight decays and regularisation [1, 9], based on findings for optimising autoencoders to prevent overfitting[1, 9]. An l2 regularisation of 0.01 was used as well as weight decay dropout of 0.2 on every layer. These were set as initial values to be experimented with but as noted by Borland et al. [1] this technique is most suited to "over-complete" autoencoders. Appearing not to be applicable here, no further experimentation was done to this values. The third model was a further expansion of the initial structure adding another additional hidden layer to the encoder/decoder phases.

### 4. DATASET

"Many-hot" encoding is based on the encoding of MIDI data. MIDI as a dataset proves significant challenges[13] in the context of the goals of accessibility of this system. Chiefly, that required quantities and qualities of MIDI data are not readily available to the average music practitioner. MIDI datasets, while available are often not consistently formatted or do not represent enough stylistic versatility as to be applicable to needs of the contemporary practitioner.

For the sake of accessibility, the use of audio sample "loops" was selected. This is due to the availability of large quantities of free and commercially available "sample packs", as well as their existing place within the musical practice of many electronic musicians and artists. With the intention that the training process be opened up to the user as in [16, 11], this guided the encoding process from audio to MIDI to "many-hot" detailed above.

The dataset used in this instance was of 168 audio loops of synthesizer sequences. These were sourced from the author's personal sample library. All loops were 120 beats-per-minute and two bars in length. Stylistically, this dataset reflects the artistic sensibilities of the author. Furthermore, in the interests of accessibility, a relatively small dataset was used. This was in order to assess whether useable results could be obtained from a corpus of material representing what an average music practitioner has access to.

### 5. SYSTEM IMPLEMENTATION

#### 5.1. Keras, Tensorflow and Python Libraries

All machine learning systems in this project were written in Python utilising the Keras and Tensorflow libraries. For the training data pre-processing and encoding the Librosa and pretty_midi libraries were utilised.

## 5.2. Pure Data & Open Sound Protocol

For the sequencer itself, the visual programming language Pure Data was used. This represents the current main user interface (Figure 2). For the steering of the AI-element there are user controls to input values for the latent vectors (X and Y) and a control to threshold the model's output. There are also conventional sequencer controls; eight steps of pitch sequencing and a length of sequence control. The sequencer's output is then sent out of the Pure Data patch as MIDI. For velocity, a separate output from the model is taken, its values scaled to between 0 and 127.

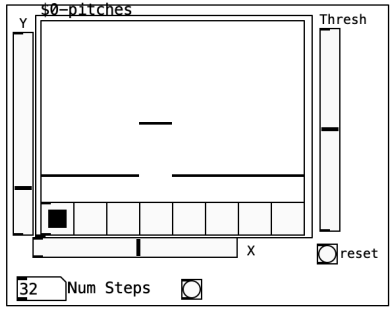

Figure 2: Sequencer user interface in Pure Data

Open Sound Control (OSC) is used to send information between the Python script and Pure Data. X,Y latent vector and threshold values are sent from Pure Data to the script, with the model constantly outputting new inferences based on the three user inputs. After every inference, the thresholded output, an array of binary values representing the generated rhythmic pattern is sent from the Python script to Pure Data, where it is unpacked and mapped to a thirty-two step-sequence.

## 6. EVALUATION & DISCUSSION

### 6.1. Methods for Evaluation

As stated, the development of this system resulted in the creation of several models, starting with the simplest, "prototype" model before being expanded to include more complex structures and techniques. After training, the encoder and decoder stages were recalled separately. The encoder stage had the training data fed into it and the latent vectors of each one plotted, thus giving a visualisation of the spread of the training data within the latent space.

The decoder was used in the generative implementation described above and trialled experientially by the author to assess its usefulness as a creative tool. In order to explore the creative uses of the system, the MIDI data from the Pure Data patch was sent to VCV rack, a free open-source emulation of the Eurorack modular synthesizer. Pitch, gate and velocity could then be patched into a system as needed.

### 6.2. Structures in Use vs. Latent Vector Visualisations

As described, the initial "prototype" architecture represented the smallest, simplest structure. Its latent vector visualisation (Figure 3) illustrates a spread of data across values between zero and four on both axes.

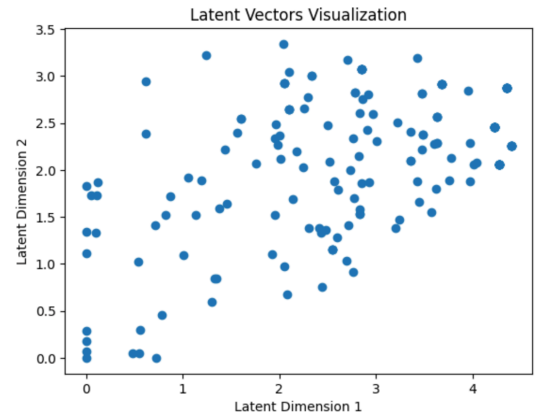

Figure 3: Latent vectors for prototype model

This model was used to establish the baseline functionality of the system including the sending of generated rhythms over OSC. In this role, the model was successful however, in use it was prone to repetition across different latent values. Its range of generative capability made it feel limited as a creative tool.

The next logical step was to expand the autoencoder structure. After some experimentation, a structure of two encoder/decoder layers of twenty and eight plus a latent layer of two was used. The trained model's latent vector (Figure 4) can be seen to have values ranging from zero to five. While there is an observable spread to the encoding, there also appears to be upward gradient of values. This is behaviour observable in every other structure as well.

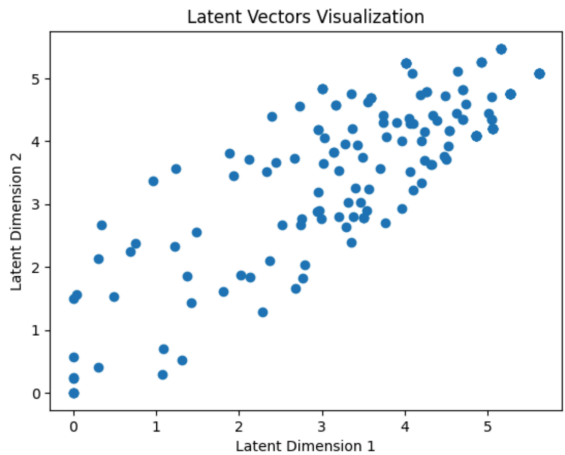

**Figure 4: Latent Vectors for Model1 with one addition hidden layer at encoder and decoder stages.**

In use, this model proved to be more creatively satisfying. Its greater range of values and greater feature extraction resulting in a much more overt range of generated rhythms.

Expanding on this model, the same structure was used with weigh decay and regularisation added to the training process. As stated, the initial intent was to prevent overfitting, however the trained latent vectors often revealed small, clustered values as in Figure 6. Intuitively, this appeared to present a less usable encoding/decoder model.

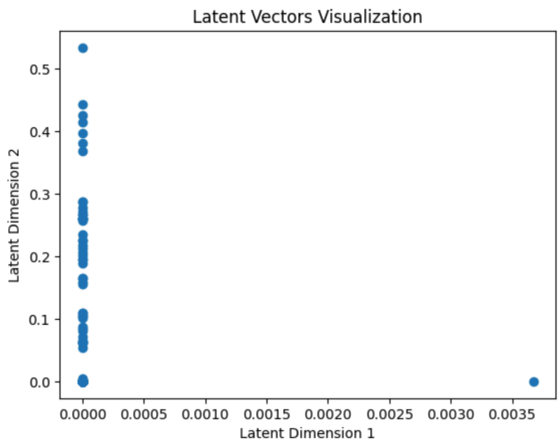

**Figure 6: Latent Vectors for Model2 with weight decay and regularisation**

In use however, this proved to be equally engaging as a creative tool. This may be due to the affordance of the user interface's values.

For both Model1 and Model2, possible X and Y values are between -10 and 10. This was initially based on the latent vectors present in the prototype and Model1, allowing for generated data beyond and the inverse of (in the case of negative values), the training data. This was to allow for the greatest possible creative output from the latent vector. In the case of Model2, though the values seemed inappropriate based on the results of the latent vector encoding, the range of results present in use was very engaging.

The final model trialled was a further expansion with encoding and decoding layers of dimensions twenty, ten and five plus the two dimensional latent layer as in previous models. As seen in Figure 6, latent values in this model ranged from zero to one but showed a relative spread of data comparable to the other models. As such, X and Y input controls were scaled to between -2 and 2. This model also yielded a large and useable range of results, displaying notably different characteristics to the others.

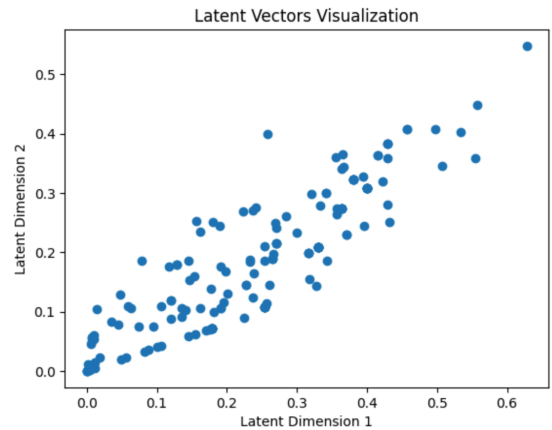

**Figure 7: Latent vectors for Model3 with added layers of encoding and decoding**

### 6.3. Overall User Evaluation

In use, all three models provided musically usable, satisfying and, importantly, distinct results. All did however, exhibit some form of notable repetitions across their generative latent vectors. The range of generated rhythms could indeed have been greater across all models. This may be attributed to the relatively small training dataset or to the low dimensionality of the latent vector. As in [16], it could be argued that superior results might be achieved by using a greater number of latent dimensions, at the expensive of the intuitive X,Y coordinate-like interface.

To counter this, affordances for the user to manipulate the sequence expanded the possibilities greatly. The ability to constrain the sequence to steps fewer than thirty-two and control over the thresholding of the models' outputs allowed for further musically satisfying results.

During evaluation, all three models were run simultaneously. In testing the models' capabilities against each other, it became evident that they represented three distinct agents. Though the same dataset had been used to train all three, notably different features had been extracted across the models. The result is that beyond seeking to find a "best" model, using the three models as an ensemble also become a viable solution. Importantly also, that the system was suitably computationally light to allow for this.

# 7. CONCLUSIONS AND FUTURE WORK

## 7.1 Conclusions

Reviewing the goals set out for the system, this implementation satisfies its requirements. The stacked autoencoder structure is able to function as a rhythm generator and the training pipeline successfully allows for the use of audio loops with a corpus size within reach of most music practitioners. Training occurs rapidly using the binary encoded data and inference is such that generation occurs at near-real-time.

Worth noting here is that for all cases, training did not provide consistent results. Each time a model was trained and its latent vectors visualised, notably different results could be seen. It was therefore necessary to train, visualise and trial several versions of the same structure, saving the most successful models for later recall.

That the system is capable of training quickly though, meant that this had the potential to become a creative element in itself. This is evident in that the inclination this author intuitively moved towards, was that of using the three trained models as distinct agents providing different "voices" in an ensemble. Though larger scale generalisation may not be achievable in such an implementation, this was not the intention. The ability to specialise and have affordances[8] for users to tailor the system to their uses was of much greater focus and the work here represents movement towards that.

## 7.2 Future Work

The work here represents several "proof-of-concepts", with further development required before the implementation would be suitable for more general use. Chiefly here would be the need for a user interface for model training, saving and recall, possibly as a stand-alone application as in [7] as well as improvements to the existing user interface.

Implementation as DAW plugin or Max for Live device as in [16, 11] would be advantageous for integration with many practitioners' existing workflows. This could also potentially be implemented as a dedicated piece of hardware to fit into the workflow of electronic musicians and those working in the Eurorack ecosystem.

In the shorter term, further experimentation with dataset sizes and contents, as well as more possible autoencoder structures could yield further potentials for neural networks and deep learning systems to function better in roles more akin to musical instrument interfaces, rather than in total AMG.